\documentclass[10pt,twocolumn,twoside]{IEEEtran}

\usepackage{amssymb}
\usepackage{enumerate}
\usepackage{graphicx,times,psfig,amsmath,amsfonts}
\usepackage{mathrsfs}
\usepackage{algorithm}
\usepackage{algorithmic}
\usepackage{multirow}
\usepackage{amsmath,amsthm}

\usepackage{lipsum}
\usepackage{subfigure}

\begin{document}
\title{A Semiblind Two-Way Training Method for Discriminatory Channel Estimation in MIMO Systems}
\author{Junjie Yang,~\IEEEmembership{}
        Shengli Xie,~\IEEEmembership{Senior Member,~ IEEE,}
        Xiangyun Zhou,~\IEEEmembership{Member,~IEEE,} 
        Rong Yu,~\IEEEmembership{Member,~IEEE,}           
        Yan Zhang,~\IEEEmembership{Senior Member,~IEEE}     
\thanks{J. Yang, S. Xie, R. Yu are with the School
of Automation, Guangdong University of Technology, Guangzhou, 510006, China.
(e-mail: yangjunjie1985@gmail.com, shlxie@gdut.edu.cn, yurong@gdut.edu.cn)} 
\thanks{X. Zhou is with the Research School of Engineering, the Australian National
University, Canberra, ACT 0200, Australia. (e-mail: xiangyun.zhou@anu.edu.au).}
\thanks{Y. Zhang is with Simula Research Laboratory, Norway; and also with Department of Informatics, University of Oslo, Norway.
(email: yanzhang@simula.no)}
}
\maketitle
\begin{abstract}
Discriminatory channel estimation (DCE) is a recently developed strategy to
enlarge the performance difference between a legitimate receiver (LR) and an
unauthorized receiver (UR) in a multiple-input multiple-output (MIMO)
wireless system. Specifically, it makes use of properly designed training
signals to degrade channel estimation at the UR which in turn limits
the UR's eavesdropping capability during data transmission. In this paper, we
propose a new two-way training scheme for DCE through exploiting a
whitening-rotation (WR) based semiblind method. To characterize the
performance of DCE, a closed-form expression of the normalized mean squared
error (NMSE) of the channel estimation is derived for both the LR and the UR.
Furthermore, the developed analytical results on NMSE are utilized to perform
optimal power allocation between the training signal and artificial noise
(AN). The advantages of our proposed DCE scheme are two folds: 1) compared to
the existing DCE scheme based on the linear minimum mean square error (LMMSE)
channel estimator, the proposed scheme adopts a semiblind approach and
 achieves better DCE performance; 2) the proposed
scheme is robust against active eavesdropping with the pilot contamination
attack, whereas the existing scheme fails under such an attack.
\end{abstract}
\IEEEpeerreviewmaketitle

\section{Introduction}

Eavesdropping by unauthorized receivers has become a prevalent security
threat in wireless communications due to the broadcast nature of the wireless
medium. Therefore, discriminating the signal reception performance between a
legitimate receiver (LR) and an unauthorized receiver (UR) becomes an
important issue in secure communications [1,2]. To address the issue, the
concept of physical layer security [3,4] has been introduced which utilizes the
physical layer properties of a wireless channel to achieve the desired
discriminatory channel performance. From an information-theoretic perspective, the
studies in [5-7] showed that the maximal data rate can be achieved by
exploiting the difference in the channel conditions between the LR and the
UR, while preventing the UR from eavesdropping any information from the
received signals. Moreover, [8] investigated the secrecy improvement
resulting from frequency selectivity  in MIMO-OFDM systems.
 From a signal processing perspective, various beamforming
schemes [9,10] have been developed to enhance the signal reception at the LR
whilst limiting the quality of signal at the UR.
The beamforming design requires the channel state information (CSI) of the LR and/or the UR priori.
The physical layer security can be used in many application senarios [5-7], \emph{e.g.,} the two-way relaying [11-13].

 Several studies on physical layer security mainly focus on data transmission without the assumption of perfect CSI.
 The pilot transmission phase is the period to acquire CSI. It is known that
channel estimation performance has a significant effect on data detection.
This observation has motivated the development of a new training strategy
called discriminatory channel estimation (DCE) such that the channel
estimation at the UR is much worse than the channel estimation at the LR. For
this, artificial noise (AN) [14-16] is inserted in the training signals to jam
the UR while keeping a minimal level of interference to the LR. Chang
\textit{et al.} [17] first designed a DCE scheme by employing multiple
feedback-and-training processes. This scheme requires large training overhead
and high design complexity. Later, a two-way training based DCE scheme was
proposed in the study [18] to reduce overhead and hence improve the
efficiency of DCE over the original scheme in [17]. The two-way DCE scheme in
[18] works well against passive eavesdropping attack from the UR. However, as
we will show in this paper, the two-way DCE scheme is not able to achieve the
desired performance under an \textit{active} eavesdropping  named the
pilot contamination attack [19]. The pilot contamination attack makes use of
the fixed and publicly known training sequence used by the LR in the reverse
training phase in order to influence the channel estimation at the
transmitter (TX). Therefore, it is important to design a robust DCE scheme
against the pilot contamination attack, which does not require any fixed and
known training sequence at the LR.

In this paper, we propose a new two-way training scheme via
whitening-rotation (WR) based semiblind approach [20-22]. The proposed scheme
includes two phases: 1) the LR transmits a sequence of stochastic signals in
the reverse training phase. These signals are only known by the LR itself,
facilitating CSI acquisition at the TX; 2) the TX broadcasts a new sequence of pilots
 inserted by AN in the forward training phase. This will
enable the channel estimation at the LR while disrupting the channel
estimation at the UR. With respect to the proposed scheme, we have the
following contributions.
\begin{itemize}

\item
The proposed WR-based DCE scheme achieves a better DCE performance
 than the existing LMMSE-based DCE scheme in [18].
By combining the blind and training-based algorithms, the WR-based semiblind techniques can
potentially enhance the quality of DCE. As shown in our numerical results,
the proposed scheme outperforms the existing scheme
with the same training overhead.
\item Another advantage of the proposed DCE scheme is the provision of
a way to protect against the pilot contamination attack due to the randomness
feature in the training signal used by the LR. We present an analytical model
to demonstrate effective attack protection and show that such attack has a
minor impact on the DCE performance.
\item  Moreover, we analytically evaluate the DCE performance
 of the proposed scheme by deriving the NMSE of channel estimation at both the LR and the UR. The optimal
 power allocation between the training signals and the AN is also
 investigated and an efficient solution is obtained as an one-dimensional line search.
\end{itemize}

The remainder of this paper is organized as follows. First,
 the system model and problem description are presented in Section II.
The proposed two-way training scheme using the WR-based channel estimator is
presented in Section III. The performance analysis under the pilot
contamination attack is discussed in Section IV. Next, simulation results are
given in Section V, followed by the conclusions in Section VI. Through the
paper, we adopt the following notations:
\begin{table}[htbp]
\centering \caption{NOTATION LIST IN THIS PAPER}
\begin{tabular}{|c|c|c|c|c|}
\hline
Symbols & Notations\\
\hline
$*$ & conjugate\\
\hline
$T$  & transpose\\
\hline
$H$  & complex conjugate transpose\\
\hline
$\circ$ &  Hadamard product\\
\hline
$Tr(\cdot)$ & the trace of a matrix\\
\hline
$\parallel \ \cdot \parallel_{F}$  & Frobenius norm\\
\hline $diag(\cdot)$ & a stacking of the diagonal
 elements \\ & of the involved matrix into a vector\\
 \hline
\end{tabular}
\end{table}
\vspace{-1em}
\section{System Model and Problem description}
\subsection{System Model}
As shown in Fig.1, we consider a wireless MIMO system consisting of a
transmitter (TX), a legitimate receiver (LR) and an unauthorized receiver
(UR). In the system, the TX, the LR and the UR have $N_{T}$, $N_{L}$, and $N_{U}$
antennas ($N_{T}>N_{L}$), respectively. The TX is connected to one LR and one UR
 by means of two different communication channels, namely legitimate channel and wiretap channel.
 The legitimate channel and the wiretap channel are denoted as
$\bf{H}\in{\mathbb{C}  ^{ \emph{N}_{\emph{L}}\times \emph{N}_{\emph{T}}} }$,
$\bf{G}\in{\mathbb{C}^{\emph{N}_{\emph{U}}\times \emph{N}_{\emph{T}}}}$, respectively.
Besides, the channel from the LR to the UR is denoted as $\bf{B}\in{\mathbb{C}^{ \emph{N}_{\emph{U}}\times \emph{N}_{\emph{L}}} }$.\\
\begin{figure}[ht]
\centering
\begin{tabular}{cc}
\includegraphics [width=3.5in]{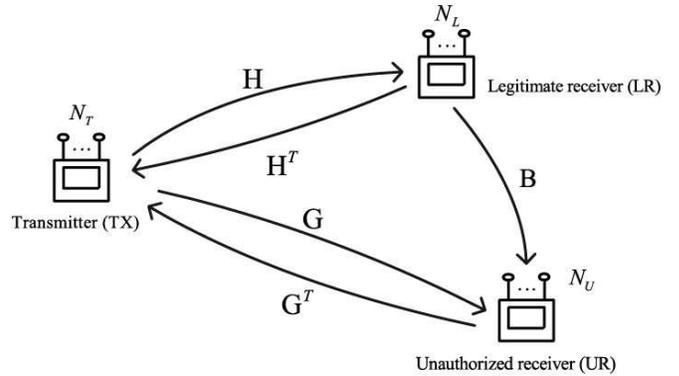}
\end{tabular}
\centering
 \caption{A wireless MIMO system includes a
multi-antennas transmitter (TX), a multi-antennas legitimate receiver (LR)
and a multi-antennas unauthorized receiver (UR).}
\end{figure}
\indent To enable the LR to interpret the legitimate channel, the TX needs to emit a sequence of training pilots, but this also
allows the UR to perform wiretap channel estimation. Therefore, the design of pilot signal is required
to ensure a high quality channel estimation at the LR but also prevent the UR from estimating the wiretap channel.
 \\
\indent The system model is based on the following assumptions.
\begin{enumerate}
\item {Channels  are assumed to be independently distributed and have reciprocity, {\emph{e.g.},} matrix $\bf{H}$ represents downlink legitimate channel and $\bf{H}$$^{{T}}$ is uplink legitimate channel. Besides, matrices $\bf{H}$, $\bf{G}$ and $\bf{B}$ are assumed to be the Rayleigh flat fading channels.}
\item { The entries of channel matrices $\bf{H}$, $\bf{G}$ and $\bf{B}$ are
 assumed to be {\emph{i.i.d},} $\mathcal C \mathcal N {(0,\sigma_{\bf{H}}^{2})}$,  $\mathcal C
\mathcal N {(0,\sigma_{\bf{G}}^{2})}$ and  $\mathcal C \mathcal N
{(0,\sigma_{\bf{B}}^{2})}$, respectively; moreover, the entries of receiver noises are  assumed to be the same as
 independent additive white Gaussian distributed, {\emph{i.i.d},} $\mathcal C \mathcal N{(0,\sigma_{0}^{2})}$}.
\end{enumerate}

The reciprocal assumption indicates that the uplink and downlink channel paths are similar, which
happens in time-division duplex (TDD) systems [23]. Such assumption is very critical for the two-way training, and this paper mainly focuses on
the design of  a DCE scheme in a TDD systems; besides, the Rayleigh flat fading channel assumption indicates that channels are relatively fixed over the transmission of symbols in one time slot but change randomly between time slots. For the second assumption of the system, it refers
that the conditions of channels and noises are respectively treated as same for the sake of fair performance.

\indent In this paper, two eavesdropping scenarios are taken into account, that is,  passive eavesdropping and active eavesdropping (pilot contamination attack [19]). For the first case, the UR only silently receives signals via channels $\bf{G}$ and $\bf{B}$ during the training phase. For the latter case, the UR not only receives signals but also emits false training pilots
 from the uplink channel $\bf{G^{\emph{T}}}$ within the training process. Such pilot contamination attack is a potential threat for the two-way training.
\vspace{-1em}
\subsection{The Existing Two-way Training Scheme}
The two-way training includes a reverse training phase and a forward training
phase. For the reverse training phase, the LR sends a reverse training signal
to the TX as
\begin{equation}
\rm{\textbf{S}}_{0} =\sqrt{\frac{\emph{P}_{0}}{\emph{N}_{\emph{L}}}\emph{T}_{0}}\bf{C}_{0}
\end{equation}
where $P_{0}$ is the power of the training pilot, and the reverse pilot
matrix $\bf{C_{0}} \in \mathbb{C}^{\emph{N}_{L}\times \emph{T}_{1}}$
satisfies an orthogonal condition
$\bf{C_{0}C_{0}^{\emph{H}}=I_{\emph{N}_{\emph{L}} }}$. The received signal at
the TX is given by

\begin{equation}
  \bf{X}_{0}=\bf{H}^{\emph{T}}\bf{S}_{0}+\bf{E}_{0},
   \label{eq:no1}
\end{equation}
where  $\bf{E}_{0}\in{\mathbb{C}^{\emph{N}_{\emph{T}}\times \emph{T}_{0}}}$
refers to the AWGN matrix. By employing the linear minimum mean-square error
(LMMSE) method [24], the channel estimation at the TX is given by
 \begin{equation}
  \bf{\hat{H}}_{0}=\sigma^{2}_{H} ( \sigma^{2}_{H}S_{0}S^{\emph{H}}_{0}+\sigma^{2}_{0}I_{\emph{N}_{\emph{L}}})^{-1}S_{0}X^{H}_{0}.
\end{equation}

In the forward training phase, the TX transmits a sequence of forward
training signals inserted with AN to enable channel estimation at the LR
while degrading the channel acquisition at the UR. In particular, the
AN-aided training sequence, denoted by $\bf{S}_{1}$, has the following
expression
\begin{equation}
\bf{S_{1}\triangleq
\sqrt{\frac{\emph{P}_{1}}{\emph{N}_{T}}\emph{T}_{1}}C_{1}}+N_{ \hat{H}_{{0}}
}A,
\end{equation}
where $P_{1}$ denotes the power of forward training pilots, $T_{1}$ is the
training length, $\bf{C_{1}}\in \mathbb{C}^{\emph{N}_{T}\times \emph{T}_{1}}$
represents the forward pilot matrix which satisfies an orthogonal condition
$\bf{C_{1}C_{1}^{\emph{H}}=I_{\emph{N}_{\emph{T}} }}$. Here, $ \bf{ N_{
\hat{H}_{{0}} }}$ is a matrix whose column vectors form an orthogonal basis
for the left null space of $\bf{\hat{H}_{{0}}}$, that is, $
\bf{N^{\emph{H}}_{ \hat{H}_{{0}} }\hat{H}_{{0}}}=0$ and $ \bf{N^{\emph{H}}_{
\hat{H}_{{0}} }N_{ \hat{H}_{{0}} } =
I_{\emph{N}_{\emph{T}}-\emph{N}_{\emph{L}}} }$. In addition, $\bf{A \in
\mathbb{C}^{(\emph{N}_{\emph{T}}-\emph{N}_{\emph{L}})\times \emph{T}_{1}}}$
is the AN matrix with each component being \emph{{i.i.d.}} $\mathcal C
\mathcal N {(0,\sigma_{a}^{2})}$.

The received signals at the LR and the UR are respectively given by
\begin{equation}
\bf{X}_{1}= \bf{H}\bf{S}_{1}+\bf{E}_{1},
\end{equation}
\begin{equation}
\bf{Y}_{1}=\bf{G}\bf{S}_{1}+\bf{F}_{1},
\label{eq:no10}
\end{equation}
where  $\bf{E}_{1} \in \mathbb{C}^{\emph{N}_{\emph{L}}\times
\emph{T}_{\bf{1}}}$ and $\bf{F}_{1}\in \mathbb{C}^{\emph{N}_{\emph{U}}\times
\emph{T}_{1}}$ are the AWGN matrices. By using the LMMSE, the channel estimation at the LR can be
expressed as
 \begin{equation}
  \bf{\hat{H}}_{1}=\sigma^{2}_{H} \{( \sigma^{2}_{H}S_{1}S^{\emph{H}}_{1}+\sigma^{2}_{0}I_{\emph{N}_{\emph{T}}})^{-1}S_{1}X^{H}_{1}\}^{\emph{T}}.
  \label{eq:no32}
\end{equation}
The UR also makes use of the received signals for its channel estimation in
the same manner as (\ref{eq:no32}), but the channel performance at the UR would be restricted due to the AN.

\section{ Proposed DCE Scheme under Passive Eavesdropping }
\subsection{ Preliminary of Whitening-Rotation based Channel Estimator }
\indent We first give a brief introduction of the WR-based semiblind channel
approach [20-22]. The channel matrix $\bf{H^{\emph{T}}}$ is firstly  estimated in the
reverse training phase, so we take the decomposition of channel
$\bf{H}^{\emph{T}}$ as an example by
\begin{equation}
\bf{H}^{\emph{T}}={W}{Q}^{\emph{H}},
 \label{eq:no002}
\end{equation}
where $\bf{W}\in {\mathbb{C}^{\emph{N}_{\emph{T}}\times \emph{N}_{\emph{L}}}
}$ is a whitening matrix and $\bf{Q}\in
{\mathbb{C}^{\emph{N}_{\emph{L}}\times \emph{N}_{\emph{L}}}}$ is an unitary
rotation matrix, \emph{i.e.},
$\bf{Q}^{\emph{H}}{Q}={Q}{Q}^{\emph{H}}={I_{\emph{N}_{\emph{L}}}}$. Besides,
performing singular value decomposition (SVD) [25] on the channel
$\bf{H^{\emph{T}}}$ gives
\begin{equation}
\bf{H}^{\emph{T}}={U}_{{H}^{\emph{T}}}\Sigma_{{H}^{\emph{T}}}
{V}_{{H}^{\emph{T}}}^{\emph{H}},
\end{equation}
where $\bf{diag(\Sigma_{H})=[\xi_{{1}},...,\xi_{{\emph{N}_{\emph{L}}}}
]^{\emph{T}}}$. One possible choice of $\bf{W}$ and $\bf{Q}$ can be
$\bf{{U}_{{H}^{\emph{T}}}\Sigma_{{H}^{\emph{T}}}} $ and
$\bf{{V}_{{H}^{\emph{T}}}}$, respectively. Without loss of generality,  the
channel estimation can be divided into two steps with the WR-based semiblind
method:
\begin{enumerate}
\item Estimate the whitening matrix $\bf{W}$ in a blind fashion using the
  autocorrelation matrix of the received signals along with a subspace based method.
\item Estimate the unitary rotation matrix $\bf{Q}$ using the training pilots with the
 constrained maximum likelihood (ML)-based method.
 \end{enumerate}

 \indent The two-steps of the WR-based channel estimator
provides a new training design for improving the DCE performance.
Specifically, the WR-based semiblind method can be used for
 the channel estimation both at the LR and the UR during the two-way training.
\subsection{Our Proposed WR-based DCE Scheme }
\subsubsection{Step I. reverse training phase}
The LR sends the reverse training signals to the TX for the uplink channel estimation
 without benefiting the channel estimation process at the UR.
In the proposed scheme, the design of reverse training signals has the same expression of (1).
Different from the existing two-way training, these reverse training signals are
 randomly generated at the LR and only known by itself, therefore,
 the TX can not apply the LMMSE method  for the channel estimation.\\
 \indent Here, the TX can resort to the blind part of
the WR-based semiblind method for the partial acquisition of channel
$\bf{H}^{\emph{T}}$. Specifically, we estimate the whitening matrix of
$\bf{H}$$^{T}$ by performing SVD on the autocorrelation matrix of the
received signals, which has the following form
\begin {equation}
\bf{{R}}_{\bf{X}_{0}}\triangleq
\frac{\bf{X}_{0}\bf{X}_{0}^{\emph{H}}}{\frac{\emph{P}_{0}}{\emph{N}_{\emph{L}}}\emph{T}_{0}}.
\label{eq:no2}
\end{equation}
\indent Referring to (\ref{eq:no2}), we can estimate the whitening matrix of $\bf{H}$$^{T}$ as
\begin{equation}
\bf{\hat{W}}_{0}=\frac{1}{\sqrt{\emph{P}_{0}}}U_{{X}_{0}}\Sigma^{\frac{1}{2}}_{{X}_{0}}
\end{equation}
by performing SVD on $\bf{R_{X_{0}}}$. By using ~(\ref{eq:no1}) ,
the autocorrelation matrix $\bf{{R}}_{X_{0}}$ can be expressed by
\begin{equation}
 \bf{ R}_{X_{0}}= H^{\emph{T}}H^{*}+ \Delta R_{X_{0}}.
 \label{eq:no001}
\end{equation}
From (\ref{eq:no001}), the error of  $\bf{R}_{X_{0}}$ has the following form
\begin{equation}
\bf{\Delta {R}_{X_{0}}=\frac{\emph{N}_{\emph{L}}}{\emph{P}_{0}}(H^{\emph{T}}\Delta {R}_{S_{0},E_{0}}+ \Delta {R}^{\emph{H}}_{S_{0},E_{0}}H^{*}+\Delta {R}_{E_{0},E_{0}}) },
\label{eq:no3}
\end{equation}
where the cross correlation matrices $\bf{\Delta {R}_{S_{0},E_{0}}}$,$\bf{\Delta {R}_{E_{0},E_{0}}}$ are defined as follows, respectively
\begin{eqnarray*}
\begin{split}
\bf{\Delta {R}_{S_{0},E_{0}}}\triangleq
\frac{S_{0}E^{\emph{H}}_{0}}{\emph{T}_{0}},\\
\bf{\Delta {R}_{E_{0},E_{0}}}\triangleq
\frac{E_{0}E^{\emph{H}}_{0}}{\emph{T}_{0}}.
\end{split}
\end{eqnarray*}

Considering the noise interference,  the error of the estimated
$\bf{\hat{W}_{0}}$ can be defined as
\begin{eqnarray}
\bf{\Delta W_{0}}\triangleq \hat{W}_{0}-W.
\end{eqnarray}
By using the results of (\ref{eq:no002}) and (\ref{eq:no3}), $\bf{\Delta{W}_{0}}$ can be deduced as
\begin{equation}
 \bf{\Delta W _{0}} = \frac{\emph{N}_{\emph{L}}}{\emph{P}_{0}}\Delta
 R_{S_{0},{E}_{0}}^{\emph{H}}Q.
\end{equation}

\subsubsection{Step II. forward training phase}

For the forward training phase, the design of forward training sequence is
required to enable the LR to interpret the downlink channel information but
degrade the channel performance at the UR. Here, the new forward training
signal is given by
\begin{equation}
\bf{S_{1}\triangleq
\sqrt{\frac{\emph{P}_{1}}{\emph{N}_{T}}\emph{T}_{1}}C_{1}} +N_{\hat{W}_{0}}A,
 \label{eq:no4}
\end{equation}
where $\bf{N}_{\hat{W}_{0}}\in \mathbb{C}^{\emph{N}_{L}\times
(\emph{N}_{T}-\emph{N}_{L})}$ is the orthogonal complement space matrix of
$\bf{\hat{W}_{0}}$ satisfying $\bf{N_{\hat{W}_{0}}^{\emph{H}}\hat{W}_{0}=0}$
and $\bf{N_{\hat{W}_{0}}^{\emph{H}}N_{\hat{W}_{0}}=I_{\emph{N}_{\emph{T}}-\emph{N}_{\emph{L}}}}$.
Compared to (4), (\ref{eq:no4})  utilizes the left null space of $\bf{\hat{W}_{0}}$
instead of
 $\bf{\hat{H}_{0}^{\emph{T}}}$ for the generation of AN. In fact,
 the left null spaces to the matrices  $\bf{{\hat{W}}_{0}}$ and $\bf{{\hat{H}}^{\emph{T}}_{0}}$ are same
 because the matrix $\bf{{\hat{W}}_{0}}$ has all the eigenvalues of $\bf{{\hat{H}}^{\emph{T}}_{0}}$.
We denote the first term of (16) as
\begin{eqnarray*}
 \bf{\tilde{S}_{1}=\sqrt{\frac{\emph{P}_{1}}{\emph{N}_{T}}\emph{T}_{1}}C_{1}}.
 \end{eqnarray*}

\indent The LR may suffer from the the imperfect estimation
of $\bf{{W}_{0}}$ due to the interference of AN, thus the power allocation problem between $P_{1}$ and $\sigma^{2}_{a}$ needs to be allocated carefully.\\

$\emph{2.1)}$ $\textbf{ \emph{Channel estimation at the LR}:}$
 Using (5) and (16), the received signal matrix at the LR can be rewritten as follows
\begin{equation}
 \bf{  LR: X_{1}=H\tilde{S}_{1}+\tilde{E}_{1}},\\
\end{equation}
where
\begin{eqnarray*}
\bf{\tilde{E}_{1}\triangleq HN_{\hat{W}_{0}}A+E_{1}}.
 \end{eqnarray*}
We apply the WR-based semiblind channel estimator for the channel estimation
at the LR. First, the whitening matrix of $\bf{H}$ can be estimated as
\begin{equation}
\bf{ {\hat{W}_{1}}=V_{\hat{X}_{W}}^{*}\Sigma^{\emph{T}}_{\hat{X}_{W}}}
\end{equation}
by performing SVD on the matrix
\begin{equation}
\bf{ \hat{X}_{W}\triangleq
\frac{X_{1}\tilde{S}_{1}^{\emph{H}}}{\frac{\emph{P}_{1}}{\emph{N}_{T}}\emph{T}_{1}}},
\end{equation}
where
\begin{eqnarray*}
 \bf{ \hat{X}_{W}=U_{\hat{X}_{W}}\Sigma_{\hat{X}_{W}}V_{\hat{X}_{W}}^{\emph{H}}}.
 \end{eqnarray*}
\indent After that, the unitary rotation matrix of $\bf{H}$ can be obtained
by solving the following optimization problem under the perturbation-free
case,
\begin {equation}
\begin {split}
&   LR: \ \ min f(\bf{Q})=\bf{
\sum\limits_{\emph{i}=1}^{\emph{N}_{\emph{L}}}\parallel X_{1}(\emph{i})-
\sum\limits_{\emph{j}=1}^{\emph{N}_{\emph{T}}}\hat{\sigma}_{\emph{j}}\emph{q}_{\emph{ij}}^{*}
\hat{S}_{1}(\emph{j})
\parallel^{2}_{\emph{F}}}\\
& \bf{s.t.\indent QQ^{\emph{H}}=I_{\emph{N}_{\emph{L}}}},
 \end {split}
 \label {eq:no23}
\end {equation}\\
 where $\bf{\hat{S}_{1}=V^{\emph{H}}_{X_{W}}\tilde{S}_{1}}$, $\bf{X_{1}(\emph{i})}$ represents the \emph{$i$}th column of $\bf{X_{1}}$ and $\bf{\hat{S}_{1}}(\emph{j})$ refers to the \emph{$j$}th column of $\bf{\hat{S}_{1}}$, respectively.
  By using the Lagrange method (the derivation can be found in appendix A),  the unitary rotation matrix can be calculated as follows
\begin {equation}
\bf{\hat{Q}_{1}=U_{\hat{X}_{Q}}V_{\hat{X}_{Q}}^{\emph{H}}}
\end {equation}
 by performing SVD on the matrix
\begin{equation}
\bf{\hat{X}_{Q}\triangleq
\frac{X_{1}^{*}\tilde{S}_{1}^{\emph{T}}\hat{W}_{1}}{\frac{\emph{P}_{1}}{\emph{N}_{\emph{T}}}\emph{T}_{1}}},
 \label{eq:no6}
\end{equation}
where
\begin{eqnarray*}
\bf{\hat{X}_{Q}=U_{\hat{X}_{Q}}\Sigma_{\hat{X}_{Q}}V_{\hat{X}_{Q}}^{\emph{H}}}.
\end{eqnarray*}
\indent According to (8), the legitimate channel $\bf{H}$ can be calculated by $\bf{\hat{H}_{1}=\hat{Q}_{1}^{\texttt{*}}\hat{W}^{\emph{T}}_{1}}$.\\

$\emph{2.2)}$ $\textbf{ \emph{Channel estimation at UR}:}$ Likewise, the
wiretape channel $\bf{G}$ can be decomposed as
\begin{equation}
\bf{G=MR^{\emph{H}}},
\end{equation}
where  $\bf{M}\in {\mathbb{C}^{\emph{N}_{\emph{U}}\times \emph{N}_{\emph{T}}}
}$ is the whitening matrix, and $\bf{R}\in
{\mathbb{C}^{\emph{N}_{\emph{T}}\times \emph{N}_{\emph{T}}} }$ is the unitary
rotation matrix. Besides, $\bf{G}$ can be decomposed by SVD as follows
\begin{equation}
 \bf{G=U_{G}\Sigma_{G}V^{\emph{H}}_{G}},
 \end{equation}
where $\bf{diag(\Sigma_{G})=[\gamma_{{1}},...,\gamma_{{\emph{N}_{\emph{U}}}}
]^{\emph{T}}}$. Without loss of generality, we can assume
that $\bf{M=U_{G}\Sigma_{G}}$ and $\bf{R=V_{G}}$.

When the UR employs the WR-based semiblind channel estimator for its channel
estimation, the received signal matrix of (6) can be rewritten as follows
\begin{equation}
\bf{ UR: Y_{1}=G\tilde{S}_{1}+\tilde{F}_{1}}, \\
\end{equation}
where $\bf{\tilde{F}_{1}\triangleq GN_{\hat{W}_{0}}A+F_{1}}$. Then, the
whitening matrix of $\bf{G}$ can be obtained as
 \begin{equation}
\bf{\hat{M}=U_{\hat{Y}_{M}}\Sigma_{\hat{Y}_{M}}}
 \end{equation}
by performing SVD to the matrix
\begin{equation}
\bf{\hat{Y}_{M}\triangleq
\frac{Y_{1}\tilde{S}_{1}^{\emph{H}}}{\frac{\emph{P}_{1}}{\emph{N}_{\emph{T}}}\emph{T}_{1}}},
\end{equation}
where
\begin{eqnarray*}
 \bf{\hat{Y}_{M}=U_{\hat{Y}_{M}}\Sigma_{\hat{Y}_{M}}V_{\hat{Y}_{M}}^{\emph{H}}}.
 \end{eqnarray*}
\indent  Next, the rotation matrix of $\bf{G}$ can be calculated using the
training-based method [22],
 \begin{equation}
\bf{ \hat{R}=V_{\hat{Y}_{R}}U_{\hat{Y}_{R}}^{\emph{H}}}
\end{equation}
by performing SVD to the matrix
 \begin{equation}
\bf{\hat{Y}_{R}\triangleq
\frac{\hat{M}^{\emph{H}}Y_{1}\tilde{S}_{1}^{\emph{H}}}{\frac{\emph{P}_{1}}{\emph{N}_{\emph{T}}}\emph{T}_{1}}},
 \label{eq:no11}
\end{equation}
where
\begin{eqnarray*}
 \bf{\hat{Y}_{R}=U_{\hat{Y}_{R}}\Sigma_{\hat{Y}_{R}}V_{\hat{Y}_{R}}^{\emph{H}}}.
\end{eqnarray*}
Using (26) and (28), the wiretap channel $\bf{G}$ can be calculated by
$\bf{\hat{G}=\hat{M}\hat{R}^{\emph{H}}}$.

\subsection{DCE Performance Analysis}
We mainly discuss the DCE performance in this Section.
A quantitative analysis of power allocation is offered for the optimal DCE scheme.

\subsubsection{Channel estimation performance at the LR}
\indent To analyze channel estimation performance, we define the perturbation
errors of $\bf{\hat{W}_{1}}$ and $\bf{\hat{Q}_{1}}$, \emph{i.e.}, $\bf{
\Delta {W}_{1}}\triangleq \hat{W}_{1}-W$ and $\bf{ \Delta {Q}_{1}}\triangleq
\hat{Q}_{1}-Q$. The estimation error of $\bf{H}$ at the LR is given by
 \begin{equation}
\bf{ \Delta H_{1}} \triangleq \hat{H}_{1}-H =\hat{Q}^{*}
\hat{W}^{\emph{T}}_{1}- Q^{*} W^{\emph{T}}_{1} \approx Q^{*}\Delta
W^{\emph{T}}_{1}+\Delta Q_{1}^{*}W^{\emph{T}}.
 \end{equation}
  \indent First, we derive the closed-form expression of $\bf{\Delta {{W}_{1}}}$. Similar to (15),
 $\bf{ \Delta {W}_{1}}$ has the following expression
\begin{equation}
\bf{\Delta
{{W}_{1}}=\frac{\emph{N}_{\emph{T}}}{\emph{P}_{1}}Q^{\emph{T}}\Delta
R^{*}_{\tilde{S}_{1},\tilde{E}_{1}}},
 \label{eq:no5}
\end{equation}
where
\begin{eqnarray*}
\begin{split}
\bf{ \Delta R_{\tilde{S}_{1},\tilde{E}_{1}}=\Delta R_{\tilde{S}_{1},E_{1}}+
\Delta R_{\tilde{S}_{1},A}  N^{\emph{{H}}}_{\hat{W}_{0}}H^{\emph{H}}}.
\end{split}
\end{eqnarray*}
 Notice that $\bf{N^{\emph{T}}_{\hat{W}_{0}}\hat{W}_{0}=0}$, (\ref{eq:no5})  can be rewritten as
\begin{equation}
\begin{split}
\bf{\Delta {{W}_{1}}=\frac{\emph{N}_{\emph{T}}}{\emph{P}_{1}}Q^{\emph{T}}
(\Delta R_{\tilde{S}_{1},E_{1}}^{*}-\frac{1}{\emph{P}_{0}}\Delta
R_{\tilde{S}_{1},A}^{*}N^{\emph{T}}_{\hat{W}_{0}}\Delta
R^{\emph{H}}_{S_{0},E_{0}})}.
\end{split}
\end{equation}
\indent  Next, we deduce the closed-form expression of $\bf{\Delta Q_{1}}$.
 Using the result of [22], $\bf{\Delta {Q}_{1}}$ equals to
\begin{equation}
\bf{\Delta{Q}_{1}\approx Q(\Gamma_{Q}\circ \Pi_{Q})^{\emph{H}}},
\label{eq:no9}
\end{equation}
where
\begin{equation}
\begin{split}
\bf{\Gamma_{Q}}=\begin{bmatrix}
\frac{1}{2{\xi}_{1}^{2}}, &...  &,\frac{1}{{\xi}_{1}^{2}+{\xi}_{\emph{N}_{\emph{L}}}^{2}} & \\
\frac{1}{{\xi}_{2}^{2}+{\xi}_{1}^{2}}, & ...&,\frac{1}{{\xi}_{2}^{2}+{\xi}_{\emph{N}_{\emph{L}}}^{2}} \\
 &.... &\\
\frac{1}{{\xi}_{{\emph{N}_{\emph{L}}}}^{2}+{\xi}_{{1}}^{2}}, &...
&,\frac{1}{2{\xi}_{{\emph{N}_{\emph{L}}}}^{2}} &
\end{bmatrix}
\end{split}
\end{equation}
and
\begin {equation}
\begin{split}
& \bf{\Pi_{Q} =\Delta X_{Q}^{\emph{H}}Q-Q^{\emph{H}}\Delta X_{Q}}.
 \label{eq:no7}
\end {split}
\end {equation}
To derive the perturbation error of $\bf{\Delta X_{Q}}$,  we define $\bf{
\hat{X}_{Q} \triangleq X_{Q}+ \Delta X_{Q}}$. Using (5), (\ref{eq:no6})  can be modified
as
\begin {equation}
\begin{split}
&\bf{ \hat{X}_{Q} = H^{*}W+H^{*}\Delta W_{1}+
\frac{\emph{N}_{\emph{T}}}{\emph{P}_{1}} \Delta
R^{\emph{T}}_{\tilde{S}_{1},\tilde{E}_{1}}W}.
\end{split}
\end {equation}
 Therefore, we have
\begin{equation}
\bf{ \Delta X_{Q}=H^{*}\Delta W_{1}+\frac{\emph{N}_{\emph{T}}}{\emph{P}_{1}}
\Delta R^{\emph{T}}_{\tilde{S}_{1},\tilde{E}_{1}}W }.
\label{eq:no8}
\end{equation}
 Substituting (\ref{eq:no8})  into (\ref{eq:no7}) , we have
\begin {equation}
\begin{split}
& \bf{\Pi_{Q}}=0.
\end {split}
\end {equation}
 As a result, (\ref{eq:no9})  can be summarized as $\bf{\Delta Q_{1}} =0$, and
 the perturbation error of $\bf{\hat{H}_{1}}$ is given by
\begin{equation}
\begin{split}
&\bf{ \Delta H_{1} = \frac{\emph{P}_{1}}{\emph{N}_{\emph{T}}} (\Delta
R_{\tilde{S}_{1},E_{1}}^{*}-\frac{\emph{N}_{\emph{L}}}{\emph{P}_{0}}\Delta
R_{\tilde{S}_{1},A}^{*}N^{\emph{T}}_{\hat{W}_{0}}\Delta
R^{\emph{H}}_{S_{0},E_{0}})}.
\end {split}
 \end{equation}
Similar to the derivation of [22], the NMSE criterion like [24] of the estimated matrix $\bf{{H}}$ at the
LR is given by
\begin{equation}
\begin{split}
&  \bf{NMSE}_{\emph{L}}\triangleq
 \frac{\rm{Tr}( E\{\bf{\Delta H_{1}\Delta
 H_{1}^{\emph{H}}\}})}{\emph{N}_{\emph{L}}\emph{N}_{\emph{T}}}\\
&= \frac{\emph{N}_{\emph{T}}\sigma_{0}^2}{\emph{P}_{1}\emph{T}_{1}}
 +\frac{\emph{N}_{\emph{L}}(\emph{N}_{\emph{T}}-\emph{N}_{\emph{L}})\sigma^{2}_{\emph{a}}}{\emph{P}_{0}\emph{T}_{0}}\frac{\emph{N}_{\emph{T}}\sigma^{2}_{0}}{\emph{P}_{1}\emph{T}_{1}}.
\end {split}
\label{eq:no21}
 \end{equation}

\subsubsection{Channel estimation performance at the UR}

We define the perturbation errors of $\bf{\hat{M}}$ and $\bf{\hat{R}}$,
\emph{i.e.}, $\bf{ \Delta {M}}\triangleq \hat{M}-M$ and $ \bf{ \Delta
{R}}\triangleq \hat{R}-R$.
 In the following, the estimation error of channel $\bf{G}$ can be given by
 \begin{equation}
 \begin{split}
&\bf{ \Delta G} \triangleq \hat{G}-G
 = \hat{M}\hat{R}^{\emph{H}}- M R^{\emph{H}} \approx M\Delta R^{\emph{H}}+\Delta MR^{\emph{H}}.
 \label{eq:no16}
\end{split}
 \end{equation}
 \indent First, we derive the closed-form expression of $\bf{\Delta M}$.
 Similar to (15), $\bf{\Delta M}$  has the following expression
\begin{equation}
\bf{\Delta M=\frac{\emph{N}_{\emph{T}}}{\emph{P}_{1}}\Delta
R^{\emph{H}}_{\tilde{S}_{1},\tilde{F}_{1}}R},
\label{eq:no28}
\end{equation}
 where
\begin{eqnarray*}
\begin{split}
\bf{\Delta R_{\tilde{S}_{1},\tilde{F}_{1}}\triangleq \Delta
R_{\tilde{S}_{1},A} N^{\emph{H}}_{\hat{W}_{0}}G^{\emph{H}}+ \Delta
R_{\tilde{S}_{1},F_{1}}}.
\end{split}
\end{eqnarray*}
 \indent   Next, we deduce the closed-form expression of $\bf{\Delta R}$.
  Similar to (\ref{eq:no9}), the perturbation matrix of  $\bf{\hat{R}}$ is given by
\begin{equation}
\bf{\Delta{R}\approx R(\Gamma_{R}\circ \Pi_{R})},
\label{eq:no14}
\end{equation}
where $\bf{\Gamma_{R}}$ is
\begin{equation}
\begin{split}
\bf{\Gamma_{R}}=\begin{bmatrix}
\frac{1}{2{\gamma}_{1}^{2}}, &...  &,\frac{1}{{\gamma}_{1}^{2}+{\gamma}_{\emph{N}_{\emph{U}}}^{2}} & \\
\frac{1}{{\gamma}_{2}^{2}+{\gamma}_{1}^{2}}, & ...&,\frac{1}{{\gamma}_{2}^{2}+{\gamma}_{\emph{N}_{\emph{U}}}^{2}} \\
 &.... &\\
\frac{1}{{\gamma}_{{\emph{N}_{\emph{U}}}}^{2}+{\gamma}_{{1}}^{2}}, &...
&,\frac{1}{2{\gamma}_{{\emph{N}_{\emph{U}}}}^{2}} &
\end{bmatrix}
\end{split}
\end{equation}
 and
\begin{equation}
\begin{split}
& \bf{\Pi_{R}=R^{\emph{H}}\Delta Y^{\emph{H}}_{R}-\Delta Y_{R}R}.
\label{eq:no13}
\end{split}
\end{equation}
 To derive the perturbation error matrix $\bf{\Delta Y_{R}}$,  we define
$\bf{ \hat{Y}_{R} \triangleq Y_{R}+ \Delta Y_{R}}$. Using (\ref{eq:no10}), (\ref{eq:no11}) can be
approximately by
\begin{equation}
\begin{split}
&\bf{\hat{Y}_{R}  \approx M^{\emph{H}}G+\Delta
M^{\emph{H}}G+\frac{\emph{N}_{T}}{\emph{P}_{1}}M^{\emph{H}}\Delta
R^{\emph{H}}_{\tilde{S}_{1},\tilde{F}_{1}}},
\end{split}
\end{equation}
Therefore,
\begin{equation}
\begin{split}
& \bf{\Delta Y_{R} = \Delta
M^{\emph{H}}G+\frac{\emph{N}_{\emph{T}}}{\emph{P}_{1}}M^{\emph{H}}\Delta
R^{\emph{H}}_{\tilde{S}_{1},\tilde{F}_{1}}}.
\label{eq:no12}
\end{split}
\end{equation}
 Substituting (\ref{eq:no12}) into (\ref{eq:no13}), we have $\bf{ \Pi_{R}=0}$. As a result,
(\ref{eq:no14}) can be summarized as $\bf{\Delta R=0}$.
Then, the perturbation matrix of $\bf{\hat{G}}$ can be yielded as
\begin{equation}
\begin{split}
&\bf{\Delta G =
\frac{\emph{N}_{\emph{T}}}{\emph{P}_{1}}(GN_{\hat{W}_{0}}\Delta
R_{\tilde{S}_{1},A}+\Delta R_{\tilde{S}_{1},F_{1}})}.
\end{split}
\end{equation}
In consequence, the NMSE criterion of estimated $\bf{{G}}$ at the UR can be
derived by
\begin{equation}
\begin{split}
 & \bf{NMSE}_{\emph{U}}\triangleq
\frac{\rm{Tr}(E{\bf{\{\Delta{G}\Delta{G}^{\emph{H}}}\}})}{\emph{N}_{\emph{T}}\emph{N}_{\emph{U}}}\\
&=\frac{\emph{N}_{\emph{T}}\sigma_{0}^{2}+\emph{N}_{\emph{T}}(\emph{N}_{\emph{T}}-\emph{N}_{\emph{L}})\sigma_{\emph{a}}^{2}\sigma_{G}^{2}}{\emph{P}_{1}\emph{T}_{1}}.
\label{eq:no15}
\end{split}
\end{equation}

Comparing (\ref{eq:no21}) and (\ref{eq:no15}), we find that the NMSE
performance at the LR and the UR are mainly relevant to the selection of
power values among the training pilots and AN. When the power of
 training pilots becomes stronger, the precision of the channel estimation
is higher for the LR but lower for the UR. Alternatively, the estimation
precision is lower for the LR while higher for the UR when the power of
training pilots becomes lower. Therefore, a power allocation trade-off
exists between the training pilots and AN.

\subsubsection{Optimal Power Allocation between the Training Pilots and AN}
One needs to allocate the available power between the training pilots and the
AN carefully through minimizing the channel estimation error at the LR while
restricting the estimation error at the UR. We formulate the power allocation
as the following optimization problem
\begin{IEEEeqnarray}{rCl}
\label{eq:no17}
& \min\limits_{P_{0},P_{1}>0,\sigma_{a}^{2}\geqslant0} ~ \bf{NMSE}_{\emph{L}} \\
&s.t. \quad  \bf{NMSE}_{\emph{U}}\geq \gamma,\IEEEyessubnumber \\
&  ~~ \quad   P_{0} \leq P_{ave}   \IEEEyessubnumber        \\
&  P_{1}+(N_{T}-N_{L})\sigma_{a}^{2}\leq P_{ave},\IEEEyessubnumber
\end{IEEEeqnarray}
where $\gamma >0$ refers to the threshold of the UR's achievable NMSE, and
$P_{ave}$ is the total energy constraint of the training signal. We define
$x=\frac{P_{1}T_{1}}{N_{T}},y=(N_{T}-N_{L})\sigma_{a}^{2}$, $z=P_{0}$. Then,
the problem (\ref{eq:no17}) can be reformulated as
\begin{IEEEeqnarray}{rCl}
\label{eq:no18}
&  \min\limits_{x>0,y\geq 0} ~   \frac{\sigma_{0}^2}{x}
+ \frac{yN_{L}\sigma^{2}_{0}}{xz} \\
&  s.t. ~~ \frac{\sigma_{0}^{2}}{x}+
\frac{y\sigma_{G}^{2}}{x} \geq  \gamma,\IEEEyessubnumber \label{eq:no35}\\
&  ~~~ \quad   z  \leq P_{ave}, \IEEEyessubnumber  \\
&  \frac{x N_{T}}{T_{1}}+y \leq  P_{ave}.\IEEEyessubnumber
\label{eq:no27}
\end{IEEEeqnarray}
\indent  The problem (\ref{eq:no18}) is a convex optimization problem involving
three variables $(x,y,z)$. We will prove that the three-dimensional
optimization problem can be solved by a simple one-dimensional line search
[26]. (The proof can be found in the Appendix B.)
\begin{itemize}[\itshape]
  \item \textbf{Proposition 1.}
Let $\{x^{*},y^{*},z^{*}\}$ be the optimal solution to the convex
optimization problem in (\ref{eq:no18}) with the constraint that
$\frac{N_{T}\sigma^{2}_{0}}{P_{ave}T_{1}}\leq \gamma \leq
(N_{T}-N_{L})P_{ave}$. The optimal value of $x$ can be solved by the
following one-dimensional optimization problem
\begin{IEEEeqnarray}{rCl}
\label{eq:no30}
 &\min\limits_{x} ~~~~ \frac{\sigma_{0}^2}{x}  + \frac{y(x)N_{L}\sigma^{2}_{0}}{xz}  \\
&s.t. ~~~  \frac{\sigma^{2}_{0}}{\gamma} \leq x \leq
\frac{(\sigma^{2}_{G}P_{ave}+\sigma^{2}_{0})T_{1}}{\gamma
T_{1}+N_{T}\sigma^{2}_{G}},\IEEEyessubnumber
\end{IEEEeqnarray}
where
\begin{eqnarray*}
\begin{split}
&  y(x^{*})=\frac{x\gamma -\sigma^{2}_{0}}{\sigma^{2}_{G}},\\
&  z^{*}   =P_{ave}.
\end{split}
\end{eqnarray*}
The associated values of $y^{*}$,$z^{*}$ are given by $y(x^{*})$, $P_{ave}$,
respectively.
\end{itemize}
\ With the result of Proposition 1, we can construct well designed training sequences for achieving optimal DCE performance.
\vspace{-1em}
\section{Proposed DCE Scheme under the Pilot Contamination Attack}
\indent  In this Section, we analyze the performance of our
 proposed DCE scheme under the pilot contamination attack.
\subsection{The Existing Pilot Contamination Attack}
\indent Under the existing two-way training scheme, the pilot contamination attack is possible when the reverse training
pilots are known by the UR. Notice that the publicly known reverse training
pilots provide an opportunity for the UR to make an adverse
 influence on the channel estimation at the TX. In particular, the UR sends
  the reverse training pilots at the same time as the LR's transmission
  during the reverse training phase. With the additive AWGN matrix $\bf{F}_{0}\in{\mathbb{C}^{\emph{N}_{\emph{T}}\times \emph{T}_{0}}}$,
  the received signals at the TX are given by
\begin{equation}
  \bf{X}_{0}=\bf{H}^{\emph{T}}S_{0}+\bf{E}_{0}+G^{\emph{T}}\bar{S}_{0}+F_{0},
  \label{eq:no100}
\end{equation}
where
\begin{eqnarray*}
  \bf{\bar{S}_{0}=\sqrt{\frac{{\bar{\emph{P}}}_{0}}{\emph{N}_{\emph{L}}}\emph{T}_{0}}\bar{C}_{0}},
 \end{eqnarray*}
 $\bar{P}_{0}$ is the power of injected fake pilot and pilot matrix $\bar{C}_{0}$ satisfies that
$\bf{\bar{C}_{0}\bar{C}_{0}^{\emph{H}}=I_{\emph{N}_{\emph{L}} }}$. If the UR knows the
reverse training pilot, then ${{{\bar{P}}}}_{0}=P_{0}$ and ${{{\bar{C}}}}_{0}=C_{0}$. For simplicity,
we define the injected noises of (\ref{eq:no100}) as
 \begin{eqnarray}
 \bar{\bf{F}}_{0}\triangleq \bf{G}^{\emph{T}}\bar{S}_{0}+\bf{F}_{0}.
\end{eqnarray}
\indent The pilot contamination attack can be viewed as a form of malicious signal injection.
With the injection of false training pilot, the UR can degrade the TX's estimation of uplink
channel $\bf{H^{\emph{T}}}$ and also align the wiretap channel estimation for the UR.
In conclusion, the impact of the pilot contamination attack on the two-way DCE scheme has
two folds: it reduces the accuracy of the LR's estimation
of the downlink channel due to the leakage of AN;
and more seriously it increases the UR's channel performance.
\vspace{-1em}
\subsection{The Impact of Pilot Contamination Attack on the Proposed Scheme}
\indent For our proposed two-way training, the reverse training pilots are
randomly generated at the LR and only known by itself. Therefore, the
randomness feature of the reverse training pilots provides a natural way of
protecting against the pilot contamination attack. Possibly, the UR may try
to exploit the received signals from channel $\bf{B}$ via the blind
detection methods [27,28]. However, the UR would suffer from a rotation
ambiguity between the estimated channel $\bf{\hat{B}}$ and channel
$\bf{{B}}$. Therefore, it is still no use for the UR to interpret the reverse
training pilot without the cooperation of the LR. \\
\indent  Here, we consider a scenario when the UR performs pilot
contamination attack with a guessing-based way, that is,
 the pilots $\bf{\bar{ C}_{0}}$ is randomly generated by a guess way.
We employ the WR-based semiblind channel estimator for
 the DCE performance. In order to study such
 attack, we mainly focus on its impact on the NMSE performance of the estimated channel $\bf{\hat{H}_{1}}$ at the LR.
 By using (\ref{eq:no5}), the perturbation error of the whitening matrix
 to the estimated channel $\bf{\hat{H}_{1}}$ can be rewritten as
\begin{equation}
\bf{\Delta
{{W}_{1}}=\frac{\emph{N}_{\emph{T}}}{\emph{P}_{1}}Q^{\emph{T}}\Delta
R^{*}_{\tilde{S}_{1},{\tilde{E}}_{1}}},
\end{equation}
where
\begin{eqnarray*}
\begin{split}
&\bf{ \Delta R_{\tilde{S}_{1},\tilde{E}_{1}} \approx}\\
& \bf{\Delta R_{\tilde{S}_{1},E_{1}}-\frac{\emph{N}_{\emph{L}}}{\emph{P}_{0}}\Delta
R_{\tilde{S}_{1},A}N^{\emph{H}}_{\hat{W}_{0}} (\Delta
R^{\emph{T}}_{S_{0},{E}_{0}}+\Delta R^{\emph{T}}_{S_{0},\bar{F}_{0}})}.
\end{split}
\end{eqnarray*}
Similar to (\ref{eq:no9}), the perturbation error of the unitary rotation matrix can be
deduced as
 $\bf{\Delta Q_{1}} =0$.
 Hence, the perturbation error of the estimated channel $\bf{\hat{H}_{1}}$
 is given by
\begin{equation}
\begin{split}
&\bf{\Delta H_{1} } =\\
& \frac{\emph{P}_{1}}{\emph{N}_{\emph{T}}} \{\Delta
R_{\tilde{S}_{1},E_{1}}^{*}-\frac{\emph{N}_{\emph{L}}}{\emph{P}_{0}}\Delta
R_{\tilde{S}_{1},A}^{*}N^{\emph{T}}_{\hat{W}_{0}}(\Delta
R^{\emph{H}}_{S_{0},E_{0}}+\Delta R^{\emph{H}}_{S_{0},\bar{F}_{0}})\}.
\label{eq:no20}
\end {split}
 \end{equation}
With the result of (\ref{eq:no20}), the NMSE criterion of the estimated channel
$\bf{H_{1}}$ at the LR is given by
\begin{equation}
\begin{split}
 & \bf{NMSE}_{\emph{L}}\approx \frac{\emph{N}_{\emph{T}}\sigma^{2}_{0}}{\emph{P}_{1}\emph{T}_{1}}
 +\frac{\emph{N}_{\emph{L}}(\emph{N}_{\emph{T}}-\emph{N}_{\emph{L}})\sigma^{2}_{\emph{a}}}{\emph{P}_{0}\emph{T}_{0}}\frac{\emph{N}_{\emph{T}}\sigma^{2}_{0}}{\emph{P}_{1}\emph{T}_{1}}\\
&+\frac{\emph{N}_{\emph{L}}(\emph{N}_{\emph{T}}-\emph{N}_{\emph{L}})\sigma^{2}_{\emph{a}}}{\emph{P}_{1}\emph{T}_{1}}
(\frac{ \emph{N}_{\emph{T}}\sigma^{2}_{0}}{\bar{\emph{P}}_{0}\emph{T}_{0}} + \frac{\emph{N}_{\emph{U}}\sigma^{2}_{G}}{{\emph{P}}_{0}\emph{T}_{0}}).
\end {split}
\label{eq:no22}
 \end{equation}

When we compare (\ref{eq:no22}) and (\ref{eq:no21}), the third term of (\ref{eq:no22}) is newly introduced
  due to the injection of fake training pilot. The third term
is significantly smaller than the first of two terms. Therefore, the pilot
contamination attack only marginally increases the NMSE error at the LR.
Furthermore, (\ref{eq:no22}) demonstrates that the power allocation using Proposition 1
is  still valid under the pilot contamination attack.

\section{Numerical results}

We consider a MIMO wireless system with one LR and one UR. In the system, we
have $N_{T}=4, N_{L}=2$ and $N_{U}=2$. Channel matrices
$\bf{H}$ or $\bf{G}$ are \emph{ i.i.d.} complex Gaussian random variables
with zero mean and unit variance ($\bf{\sigma^{2}_{H}=\sigma^{2}_{G}=1}$); and the additive noise
matrices $\bf{E_{0}}$, $\bf{E}_{1}$ or $\bf{F}_{1}$ are \emph{i.i.d.} AWGN ($\sigma^{2}_{0}=0.01$). We set the maximum transmission power as 30 dBm,
\emph{i.e.,} $P_{ave}=1$. The reverse training pilot matrix and the forward pilot matrix
satisfy that  $\bf{C_{0}C_{0}^{\emph{H}}=I_{\emph{N}_{\emph{L}} }}$, $\bf{C_{1}C_{1}^{\emph{H}}=I_{\emph{N}_{\emph{T}} }}$, respectively.
Moreover, the overall training length is given as $T=280$, in which
$T_{0}=T_{1}=140$. The parameter $\gamma$ is set as 0.03 or 0.1 [17].
With $P_{ave}=1$, the criterion of
signal-to-noise ratios (SNRs) at the LR and the UR can be defined  as
\begin{equation}
\begin{split}
\bf{SNR}_{\emph{L}}=\frac{\rm{E}\bf{\parallel HS_{1} \parallel_{\emph{F}}^{2}}}{\rm{E}\bf{\parallel E_{1}\parallel_{\emph{F}}^{2}}}=\frac{1}{\sigma_{0}^{2}},\\
\bf{SNR}_{\emph{U}}=\bf{\frac{\rm{E}{\parallel \bf{GS}_{1} \parallel_{\emph{F}}^{2}}}{\rm{E}{\parallel \bf{F}_{1} \parallel_{\emph{F}}^{2}}}}=\frac{1}{\sigma_{0}^{2}},\\
 \end{split}
\end{equation}
where $\bf{SNR}_{\emph{L}}=\bf{SNR}_{\emph{U}}$. Here,  the DCE scheme based on the LMMSE channel estimator in [18]
is employed as a fair comparison to the proposed  DCE scheme based on the WR channel estimator. In the
 following, the scheme in [18] is called as  LMMSE-based DCE scheme, and the proposed scheme is named as WR-based DCE scheme.
 The result of each DCE scheme is obtained over 100,000 Monte Carlo running.
\begin{figure}[ht]
\centering
\begin{tabular}{cc}
\includegraphics [width=3.5in]{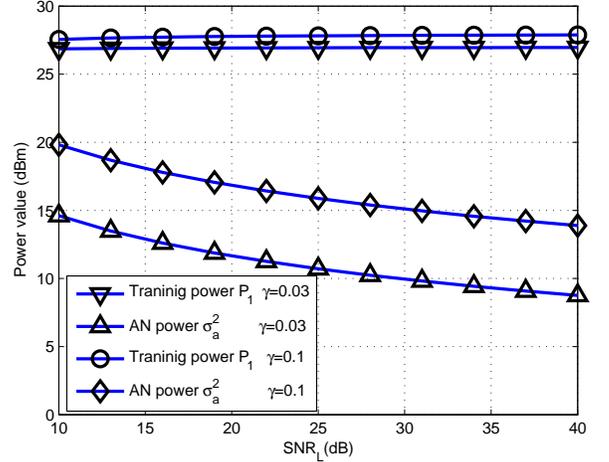}
\end{tabular}
\centering
 \caption{ Power allocation among the training signals and AN in our
  proposed DCE scheme}
\end{figure}
We first demonstrate the efficiency of the proposed DCE scheme. Both the LR and the UR exploit their channel
estimation with the received signal samples. The optimal power value among
$P_{1}$ and $\sigma^{2}_{a}$ can be calculated via the optimized solution of
(\ref{eq:no30}). Fig.2 shows that more power is needed for the forward training pilots
than the AN at all SNR levels. There is a minor variation between the powers of forward
training pilots and the AN at all SNR levels. This observation indicates
that the power allocation solution has a stable performance.\\
\indent From Fig.3 (a) and Fig.3 (b), it is found that the NMSE performance at the LR
with the WR-based DCE scheme achieves better DCE performance than the LMMSE-based DCE scheme.
By assuming that the TX knows perfect CSI of the uplink channel, we illustrate the
ideal lower bound of NMSE with the proposed DCE scheme in the figures. The
gap between the ideal lower bound with perfect CSI and the NMSE error at the LR with the LMMSE-based DCE scheme
 is wide while it is very close to the WR-based DCE scheme. Furthermore, Fig 3.(c) shows the relationship
between NMSE at the LR and the training sequence length. The NMSE performance at the LR improves the DCE performance with
longer training pilots. Although our simulation results are shown for the scenario where the LR and UR
are at the same distance from the TX ($\bf{\sigma^{2}_{H}=\sigma^{2}_{G}}$), the proposed DCE scheme works well
 even if the UR is much closer to the TX as long as the transmit power is sufficiently large.
The reason is that the SNR at the UR does not change much with distance when the inserted AN dominates the noise at the UR.
\begin{figure}
\centering \subfigure[$\gamma =0.03$ ]{
\begin{minipage}[b]{0.48\textwidth}
\includegraphics[width=1\textwidth]{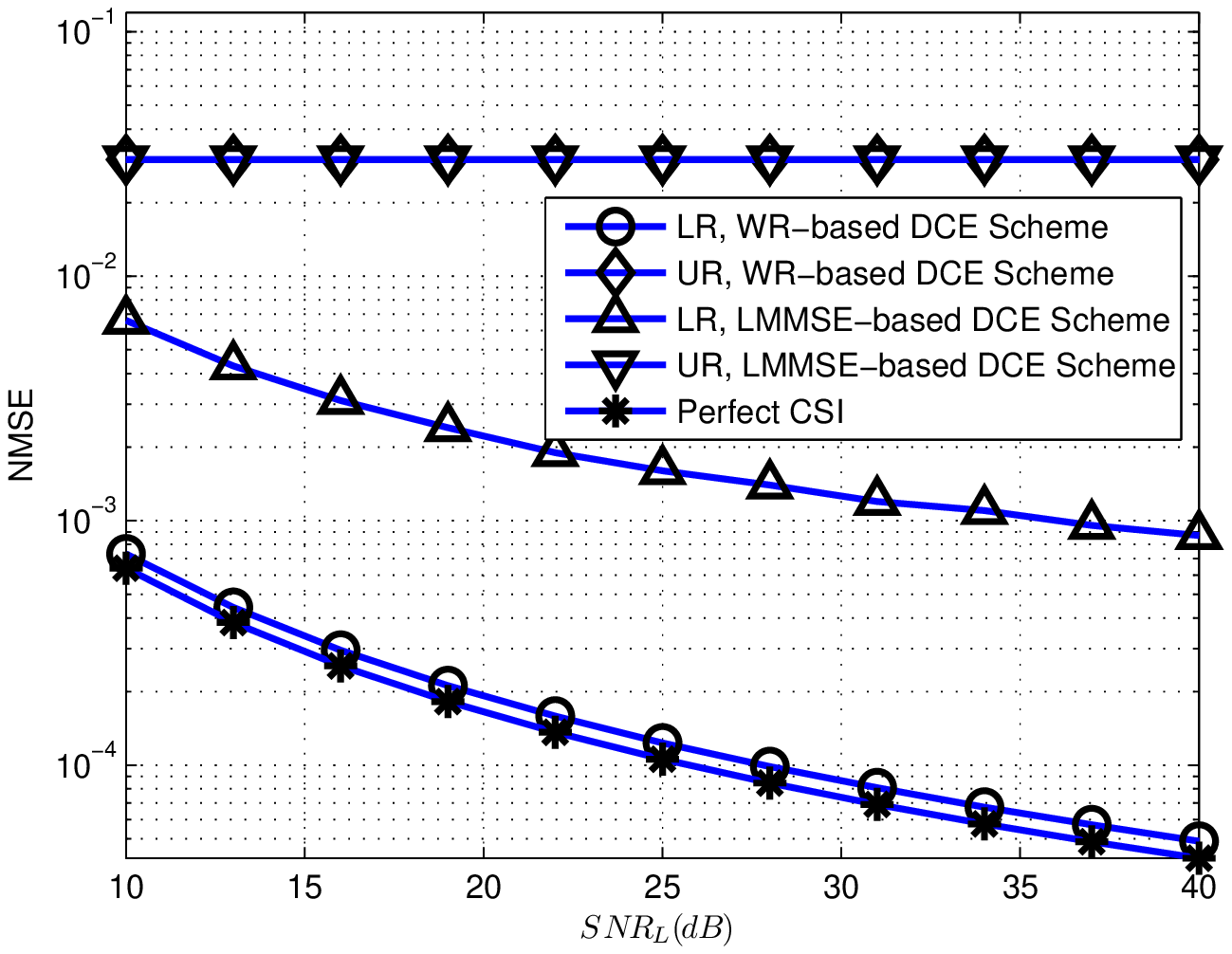} \\
\end{minipage}
} \centering \subfigure[$\gamma =0.1$]{
\begin{minipage}[b]{0.48\textwidth}
\includegraphics[width=1\textwidth]{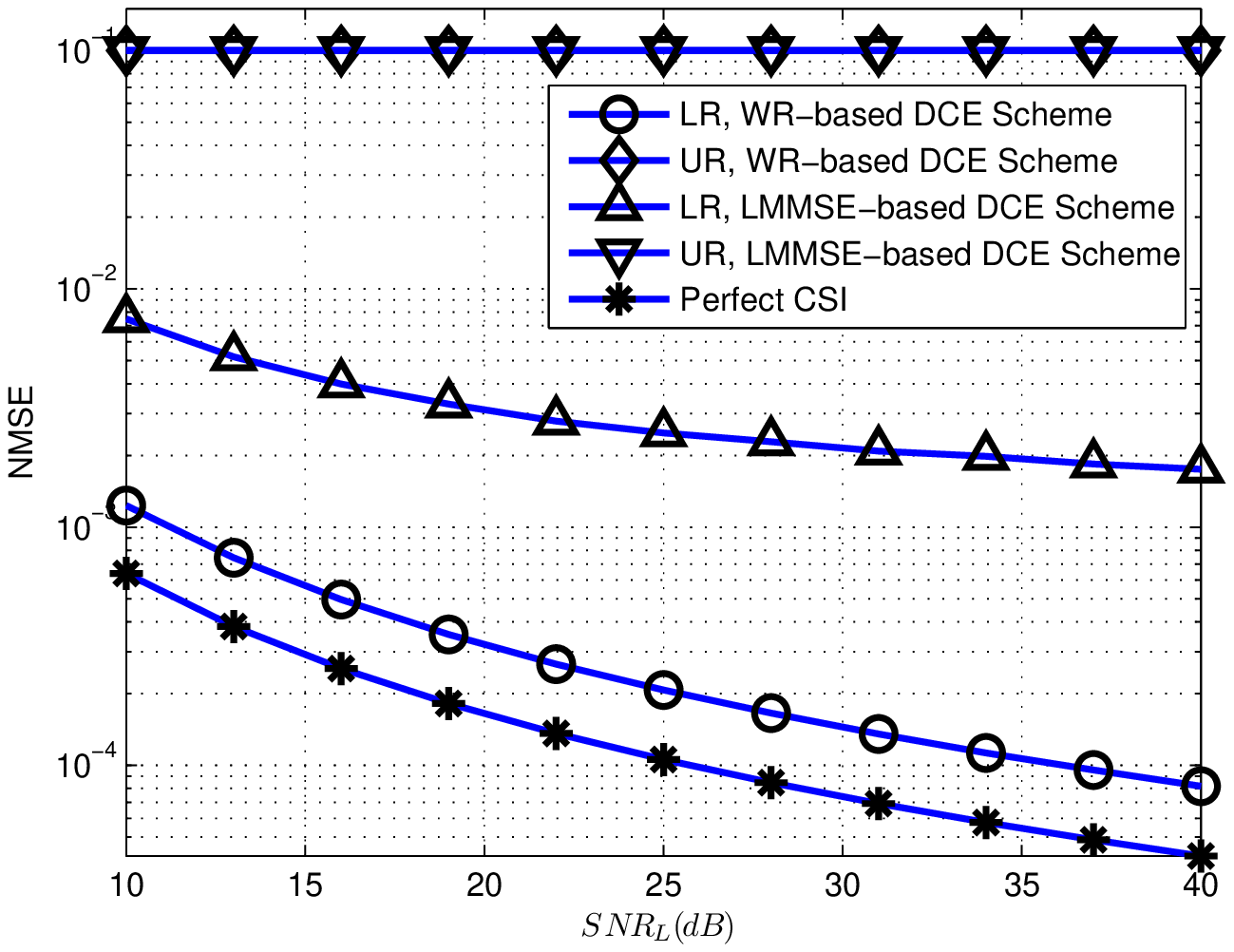} \\
\end{minipage}}
\centering \subfigure[Variation of $T_{1}$ when $SNR_{L}=25dB$.]
{\begin{minipage}[b]{0.48\textwidth}
\includegraphics[width=1\textwidth]{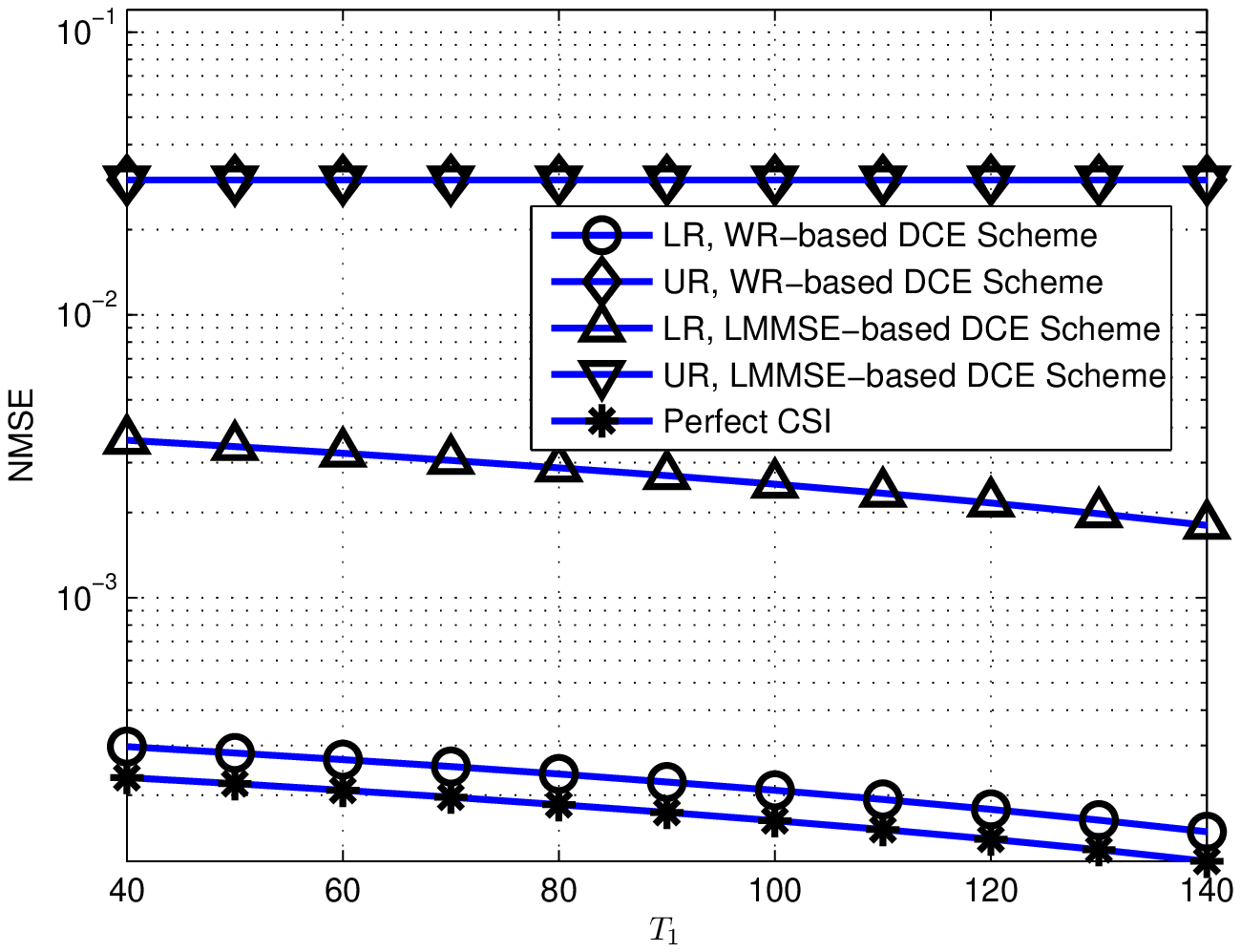} \\
\end{minipage}}
\caption{NMSE performance comparison between the LMMSE-based DCE scheme and the WR-based DCE scheme}
\end{figure}

\begin{figure}
\centering \subfigure[$\gamma =0.03$ ]{
\begin{minipage}[b]{0.48\textwidth}
\includegraphics[width=1\textwidth]{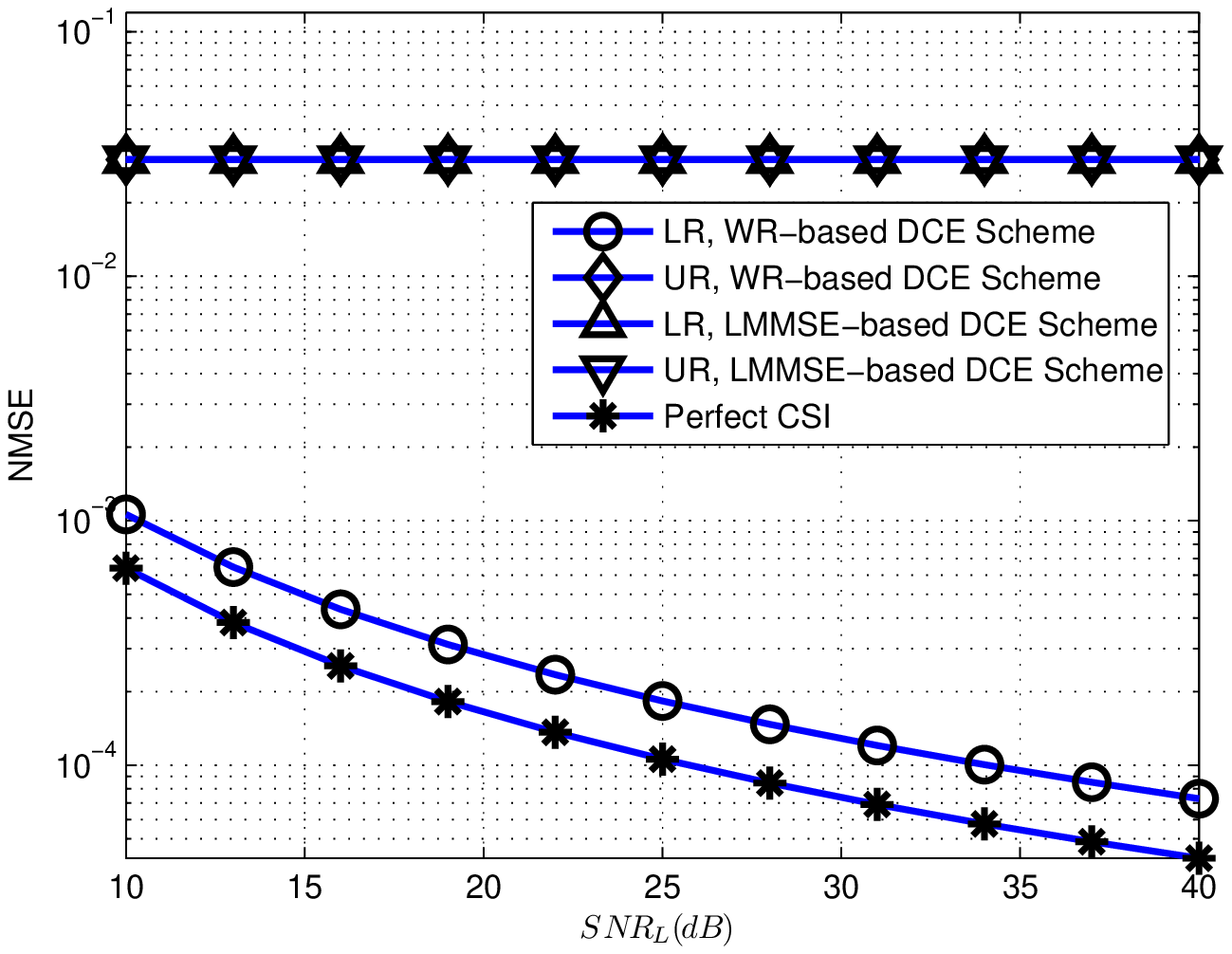} \\
\end{minipage}
} \centering \subfigure[$\gamma =0.1$]{
\begin{minipage}[b]{0.48\textwidth}
\includegraphics[width=1\textwidth]{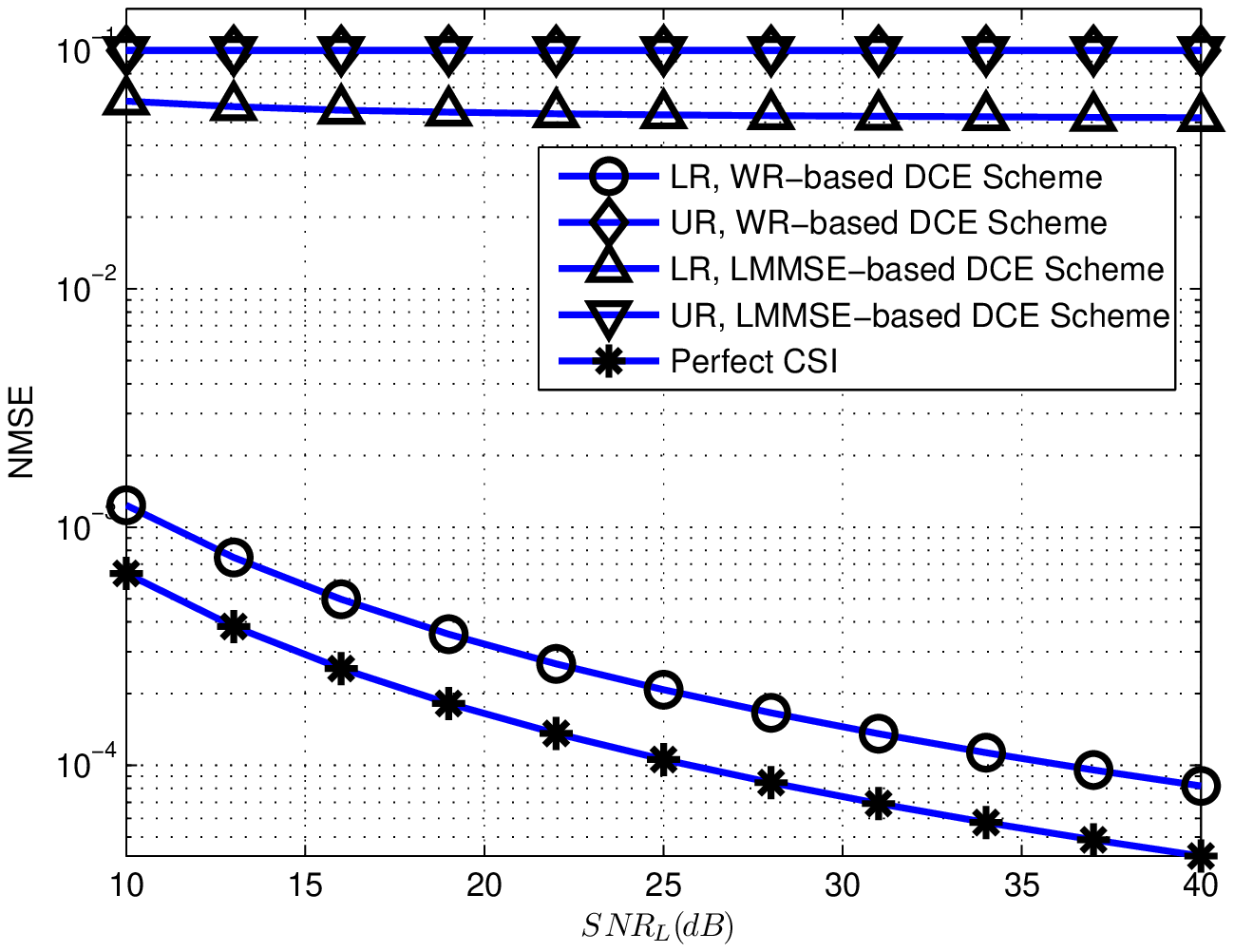} \\ 
\end{minipage}
} \centering \subfigure[Variation of $\bar{P}_{0}$ when the noise levels are
set as $15dB$, $25dB$, $35dB$, respectively.]{
\begin{minipage}[b]{0.48\textwidth}
\includegraphics[width=1\textwidth]{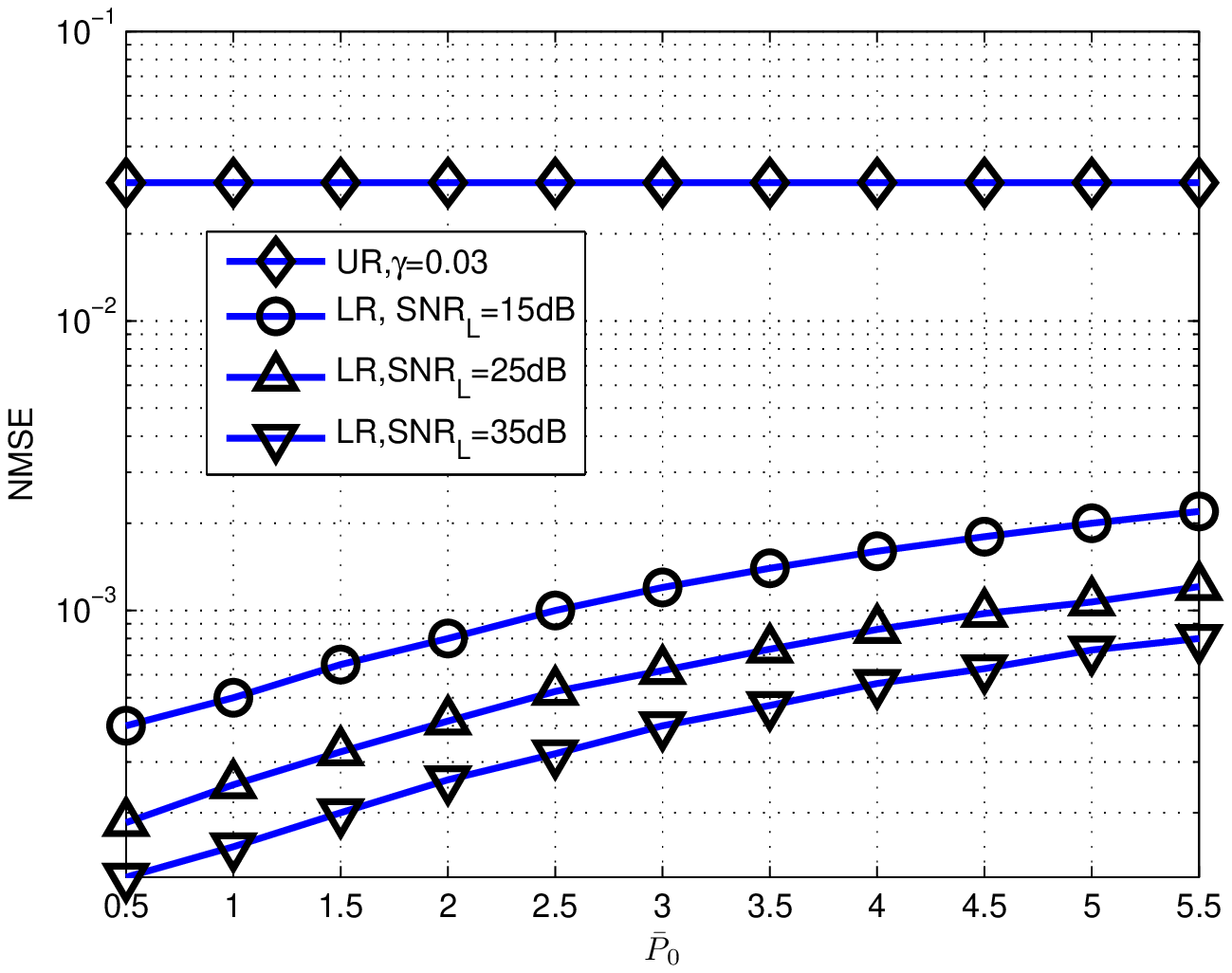} \\
\end{minipage}
} \caption{NMSE performance of the LMMSE-based DCE scheme and the WR-based DCE scheme under the pilot contamination attack}
\end{figure}

We then look into the performance of the proposed WR-based DCE scheme under a
pilot contamination attack.  As we have discussed in Section IV, the UR has
no knowledge of the reverse training signals. In addition, we suppose that the UR sends another guess-based
orthogonal signals $\bf{\bar{S}}_{0}$ to attack the system, in which $\bar{P}_{0}=P_{ave}=1$.
From Fig.4 (a), one can see that the NMSE performance
at the LR with LMMSE-based DCE scheme has approximately reached the limitation of the UR's NMSE due to the
poor channel estimation. On the contrary, the NMSE performance at the LR with
the proposed DCE scheme only has a slight reduction. Similar results can be
observed in Fig.4 (b) in case of $\gamma=0.1$.
 As a result, the LMMSE-based DCE scheme is more sensitive to
 the pilot contamination attack than our proposed DCE scheme.

Fig.4 (c) shows the impact of pilot contamination attack on the performance
of our proposed scheme with variation of attack power. Since the LMMSE-based DCE scheme
works not very well under the pilot contamination  attack, we only focus on the
performance of the proposed DCE scheme. The result show that the obtained NMSE error at the LR has
a slight reduction with the increasing power $\bar{P}_{0}$. The attack power
is restricted at the UR, hence the pilot contamination attack has an
ignorable impact on the performance of our proposed DCE scheme.
\section{Conclusion}
In this paper, we proposed a new two-way training scheme for DCE in wireless
MIMO systems. To improve the DCE performance, an efficient whitening-rotation
(WR) based semiblind approach has been employed in the two-way training. A
closed-form of NMSE on the channel estimation between the LR and the UR has
been developed, facilitating optimal power allocation between the training
signals and AN. Furthermore, the proposed training design offers a
countermeasure against the pilot contamination attack. It has been proved
that the pilot contamination attack has a very limited impact on the DCE
performance. Simulation results demonstrate that our proposed scheme can
achieve higher performance than the existing training scheme.\\
\indent Our future work includes the following directions of research:
(a) We will extend the study of DCE from narrowband systems to wideband OFDM-based systems [29].
 The power allocation between the training signal and AN across all pilot subcarriers
 will be an interesting problem to investigate. (b) We will apply the semiblind two-way training
 method to tackle the pilot contamination problem in multi-cell massive MIMO systems.
\appendices
\section{Derivation of unitary rotation matrix }
\indent According to (\ref{eq:no23}), the rotation matrix $\bf{Q}$ can be obtained by using
Lagrange method as follows,
\begin {equation}
\begin {split}
&  LR: \ \ min f(\bf{Q},\lambda,\mu)=
\sum\limits_{\emph{i}=1}^{\emph{N}_{\emph{L}}}\parallel \bf{X_{1}(\emph{i})}-
\sum\limits_{\emph{j}=1}^{\emph{N}_{\emph{T}}}\hat{\sigma}_{\emph{j}}\emph{q}_{\emph{ij}}^{*}
\bf{\hat{S}}_{1}(\emph{j})
\parallel_{\emph{F}}^{2} \\&+ \sum\limits_{i=1}^{\emph{N}_{L}} Re\{\lambda_{i}(q^{\emph{H}}_{\emph{i}}q_{i}-1)\}
+\sum\limits_{i=1}^{N_{L}}\sum\limits_{j=i+1}^{N_{L}}Re\{\mu_{ij}q^{H}_{i}q_{j}\}.
 \end {split}
 \label{eq:no24}
\end {equation}
\indent By differentiating (\ref{eq:no24}) \emph{w.r.t} $\bf{Q}$ and let it equals to
$0$, we have
\begin{equation}
\bf{X_{1}^{*}\tilde{S}_{1}^{\emph{T}}\hat{W}_{1}-Q\hat{W}_{1}^{\emph{T}}\tilde{S}_{1}\tilde{S}^{\emph{H}}_{1}\hat{W}_{1}=Q\Theta},
 \label{eq:no25}
\end{equation}
where $\Theta$ is the matrix of Lagrange multipliers which satisfies
$\Theta_{ii}=\lambda_{i}$,$\Theta_{ij}=\mu_{ij}$ when $i>j$ and
$\Theta_{ij}=\mu^{*}_{ij}$ when $i<j$.  We divide
$\frac{\emph{P}_{1}}{\emph{N}_{\emph{T}}}\emph{T}_{1}$ on both sides of (\ref{eq:no25}),
and denote
\begin{eqnarray*}
\bf{\hat{X}_{Q} \triangleq
\frac{X_{1}^{*}\hat{S}_{1}^{\emph{T}}\hat{W}_{1}}{\frac{\emph{P}_{1}}{\emph{N}_{\emph{T}}}\emph{T}_{1}}}.
\end{eqnarray*}
Using the conjugate symmetry feature of $\Theta$, we premultiply
$\bf{Q}^{\emph{H}}$ to (\ref{eq:no25}) and post-multiply  $\bf{Q}^{\emph{H}}$ to the
conjugate of (\ref{eq:no25}). Performing a subtraction between the two new equations, we
have
\begin{equation}
\bf{Q^{\emph{H}}\hat{X}_{Q}=\hat{X}_{Q}^{\emph{H}}Q}.
 \label{eq:no26}
 \end{equation}
 According to the specific structure of (\ref{eq:no26}), it holds if and only if
\begin{equation}
\begin{split}
\bf{\hat{Q}_{1}=U_{\hat{X}_{Q}}V_{\hat{X}_{Q}}^{\emph{H}}},
\end{split}
\end{equation}
where $\bf{U_{\hat{X}_{Q}}}$ and $\bf{V_{\hat{X}_{Q}}}$ are the decomposition
results after performing SVD on $\bf{\hat{X}_{Q}}$,\emph{ e.g., }
$\bf{\hat{X}_{Q}=U_{\hat{X}_{Q}}\Sigma_{\hat{X}_{Q}}V_{\hat{X}_{Q}}^{\emph{H}}}$.
\section{Derivation of optimum power allocation}
First, we need to develop the range of parameter $\gamma$. For the extreme
case when $y=0$, we know that
\begin{equation}
\centering
\begin{split}
x\leq  \frac{P_{ave}T_{1}}{N_{T}}
\end{split}
\end{equation}
 by using the power constraint of (\ref{eq:no27}). In this case,
 $\gamma$ has the lower bound
\begin{equation}
\centering
\begin{split}
 \gamma \geq \frac{N_{T}\sigma^{2}_{0}}{P_{ave}T_{1}}.
\end{split}
\end{equation}
To obtain the upper bound of $\gamma$, we need to analyze the NMSE error
performance at the UR when $x=0$. Similar to (\ref{eq:no28}), the perturbation matrix of
$\bf{M}$ can be rewritten as
\begin{equation}
\centering
\begin{split}
 \bf{ \Delta M \approx \Delta
  {R}^{\emph{H}}_{A,F_{1}}N^{\emph{H}}_{\hat{W}_{0}}R},
\end{split}
\end{equation}
 In this case, we have the result as follows
\begin{equation}
\centering
\begin{split}
 \bf{NMSE}_{\emph{U}} \approx  (\emph{N}_{\emph{T}}-\emph{N}_{\emph{L}})\emph{P}_{ave}.
\end{split}
\end{equation}
 This result implies the worst MNSE performance at the UR,  so we can obtain the upper bowed of $\gamma$,
 \begin{equation}
\gamma \leq (N_{T}-N_{L})P_{max}.
\end{equation}
\indent  For the variable $y$, it satisfies the following inequations as
\begin{equation}
\centering
\begin{split}
& \frac{x\gamma -\sigma^{2}_{0}}{\sigma^{2}_{G}} \leq y \leq P_{ave}-
\frac{xN_{T}}{T_{1}}
\end{split}
\label{eq:no29}
\end{equation}
by using the constraints of (\ref{eq:no35}) and (\ref{eq:no27}).
 Besides, the inequations of (\ref{eq:no29}) holds if and only if
\begin{equation}
\centering
\begin{split}
P_{ave}-\frac{x N_{T}}{T_{1}} \geq \frac{x\gamma
-\sigma^{2}_{0}}{\sigma^{2}_{G}}.
\end{split}
\end{equation}
Thus, $x$ satisfies
\begin{equation}
 \frac{\sigma^{2}_{0}}{\gamma} \leq x \leq
\frac{(\sigma^{2}_{G}P_{ave}+\sigma^{2}_{0})T_{1}}{\gamma
T_{1}+N_{T}\sigma^{2}_{G}}.
\end{equation}
From the optimal problem (\ref{eq:no18}), the objective function is monotonically
decreasing with respect to the variable $y$, so the optimal value can be
achieved when
\begin{equation}
  y^{*}=\frac{x\gamma -\sigma^{2}_{0}}{\sigma^{2}_{G}}.
\end{equation}
Furthermore, $z$ is independent from $x$ and $y$, thus the objective function
approaches its optimization point when $z^{*}=P_{ave}$.
\section*{Acknowledgments}
The work was supported in part by programs of NSFC under Grants nos. 61322306, 61333013, U1201253, 61273192,
61370159, U1035001, 6120311761370159, U1035001 and 61203117,
Guangdong Province Natural Science Foundation of under Grant S2011030002886 (team project),
the Department of Science and Technology of Guangdong Province, China (nos.~2011A090100039,2011B090400360),
program for New Century Excellent Talents in University under Grant NCET-11-0911 and Special Scientific Funds approved in 2011 for the Recruited Talents by Guangdong Provincial universities, and the Science and Technology Program of Guangzhou, China (grant no. 2014J2200097).
The work of X. Zhou was supported by the Australian Research Council's Discovery Projects funding scheme (grant no. DP140101133).
The corresponding author is Shengli Xie.
\end{document}